\documentstyle[lprocl,11pt,psfig]{article}

\def\Journal#1#2#3#4{{#1} {\bf #2}, #3 (#4)}

\def\NPB{{\em Nucl. Phys.} B}
\def\PLB{{\em Phys. Lett.}  B}
\def\PRL{\em Phys. Rev. Lett.}
\def\PRD{{\em Phys. Rev.} D}
\def\ZPC{{\em Z. Phys.} C}

\def\ra{\rightarrow}
\def\be{\begin{equation}}
\def\ee{\end{equation}}
\def\bea{\begin{eqnarray}}
\def\eea{\end{eqnarray}}

\newcommand{\sleq} {\raisebox{-.6ex}{${\textstyle\stackrel{<}{\sim}}$}}
\newcommand{\sgeq} {\raisebox{-.6ex}{${\textstyle\stackrel{>}{\sim}}$}}
\newcommand {\pom}  {I\hspace{-0.2em}P}
\newcommand {\reggeon}  {I\hspace{-0.2em}R}

\newcommand {\alphareg} {\mbox{$\alpha_{_{\reggeon}}$}}
\newcommand {\apom} {\mbox{$\alpha_{_{\pom}}$}}
\newcommand {\xpom} {\mbox{$x_{_{\pom}}$}}
\newcommand {\gev} {\mbox{$\mathrm{GeV}$}}
\newcommand {\gevtwo} {\mbox{$\mathrm{GeV^2}$}}
\newcommand {\gevmtwo} {\mbox{$\mathrm{GeV^{-2}}$}}
\newcommand {\ftwod} {\mbox{$F_2^{D(3)}$}}
\newcommand {\ftwodfour} {\mbox{$F_2^{D(4)}$}}

\begin{document}

\title{DIFFRACTION}

\author{ E. GALLO }
\address{INFN Firenze, Largo Enrico Fermi 2
\\ 50125 Firenze, Italy
\\ E-mail: gallo@desy.de}

%%%%%%%%%%%%%%%%%%%%%%%%%%%%%%%%%%%%%%%%%%%%%%%%%%%%%%%%%%%%%%
% You may repeat \author \address as often as necessary      
%%%%%%%%%%%%%%%%%%%%%%%%%%%%%%%%%%%%%%%%%%%%%%%%%%%%%%%%%%%%%%%
\maketitle\abstracts
{
This report summarizes recent results on diffraction obtained
at HERA, at the Tevatron and by the fixed target experiment E665. 
The measurements include vector meson production,
inclusive diffraction at HERA and the pomeron structure function as
inferred from the HERA and the Tevatron data. 
}

\section{Introduction}

Diffractive interactions have recently attracted a lot of interest,
following the results coming from HERA and the Fermilab experiments.
Let me start with a brief introduction on what is diffraction. 

Consider the total hadronic 
cross sections for the processes $pp, p \bar p, \pi^{\pm}p$,$K^{\pm} p$,$\gamma p$,
 as a function of the centre of mass (c.m.) energy
$\sqrt s$ (see e.g.~\cite{pdg}). Their behaviour  
can be described as the sum of two components:
$\sigma^{h-h}_{tot} = Y s^{-\eta} + X s^{\epsilon}$, where the
first term describes the decrease of the cross section with $s$
at low energies, the second the slow increase at high energies.
Donnachie and Landshoff~\cite{dl} have performed a fit to all the
available data at that time, obtaining a universal description of the
hadronic cross sections with values $\eta \simeq 0.45$ and 
$\epsilon \simeq 0.08$. Recently Cudell et al.\cite{cudell} have repeated
the fit using only $pp$ and $p \bar p$ data, obtaining the value
$\epsilon= 0.096^{+0.012}_{-0.009}$, but claiming that any $\epsilon$
value in the range $0.07-0.11$ could describe the data. Finally the
CDF experiment has measured $\epsilon$ with their own $p \bar p$ data
at two different values of $\sqrt s$, obtaining a somewhat higher
value, $\epsilon=0.112 \pm 0.013$ \cite{cdftot}.

This behaviour of the total hadronic cross section can be interpreted
in terms of Regge theory \cite{collins}. 
The hadronic reaction $A + B  \rightarrow C + D$ can
be described by the exchange in the $t$-channel of a family of particles,
such that the relevant quantum numbers are conserved. For these
particles there is a linear relation between the spin $J$ and the mass
squared ($t$), which is of the form 
$J=\alpha(t) = \alpha(0)+ \alpha^\prime t$.
The particles are lying on a so-called Regge `trajectory', with intercept
$\alpha(0)$ and slope $\alpha^\prime$. 
Regge theory predicts that the total cross section should behave as:
\begin{equation}
\sigma_{tot} \propto s^{\alpha(0)-1}.
\end{equation}
The dependence of the elastic cross section with $t$, which is found
to be exponential at small values of $t$, should behave as:
\begin{equation}
\frac {d \sigma_{el}}{dt} \propto (\frac{s}{s_0})^{2(\alpha(0)-1)} e^{bt},~~~
\mathrm{with}~~~b = b_0 + 2 \alpha^\prime \ln(s/s_0)
\end{equation}

The slowly decreasing term at low energy fitted by Donnachie-Landshoff
corresponds to the intercept $\alphareg \simeq 0.5$ for reggeon exchange
(i.e. the degenerate $\rho,\omega,f$ and $a$ trajectories).
The slow increase of the cross section at large energies corresponds
to the so called pomeron trajectory, which has an intercept $\simeq 1.08$
according to the fit in \cite{dl}, and which is the
dominating term at high energies. The value of $\alpha^\prime_{\pom}$ 
was fitted to be $\simeq 0.25~\gevmtwo$, implying
that the exponential $t$ distribution becomes steeper as the energy increases
(an effect called `shrinkage'). 
There are no known particles corresponding
to the pomeron, except a glueball candidate 
from the WA91 experiment \cite{wa91} which would lie on its trajectory.

The $d\sigma/dt$ distribution in proton-proton elastic scattering has a characteristic
behaviour, with an exponential fall-off, a dip and a second exponential,
which is very similar to the pattern of diffraction of light by a circular
aperture (see e.g.~\cite{goulianos}). Therefore the name of {\em Diffraction}
was used to indicate pomeron exchange.

\begin{figure}
\begin{center}
\mbox{
\psfig{figure=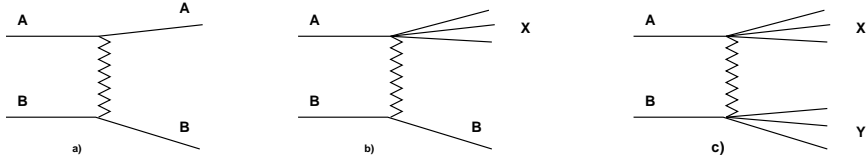,height=2cm,angle=-90}}
\end{center}
\caption{Feynman diagrams corresponding to pomeron exchange: 
( a) elastic, b) single dissociation, c) double
dissociation.
\label{fig:diag}}
\end{figure}

The processes mediated by pomeron exchange can be classified as
elastic, single dissociation and double dissociation as shown is 
fig.~\ref{fig:diag}. In this report I will concentrate on elastic
vector meson production and the single dissociation process.
In the latter process, the particle 'B' is either a proton or anti-proton,
which remains intact after the interaction, carrying almost all of the initial
beam momentum; the particle A can be either a proton or antiproton
at the Tevatron, or a photon or virtual photon at HERA and E665. In Vector Meson
Dominance (VDM) models or in perturbative QCD (pQCD), the photon can
fluctuate into hadrons, so that the $\gamma^{(\ast)}p$ interaction can
be seen as a hadron-hadron interaction. From the processes pictured
in fig.~\ref{fig:diag} it is also clear that the pomeron has to carry
the quantum numbers of the vacuum: in particular it is a colour-singlet and one
expects to see a rapidity gap (i.e. a region with no particles)
 between the leading particle B and the dissociative system $X$.  
The experiments at HERA and the Tevatron have large rapidity coverage and
have or are planning to have a forward proton spectrometer 
to measure the scattered leading proton.

There are still many open questions in diffraction on the nature of the
pomeron. One of the main issues is whether the pomeron is `soft', by which it is
generally meant that
its intercept is close to 1.08, or whether is it `hard', which corresponds
to an effective intercept greater than 1.08. In pQCD models~\cite{workshop},
the dependence of the cross section with the energy is driven by the rise
of the gluon density at low $x$,  where $x$ is the Bjorken variable, and a
steep increase with the energy may be expected.
This will be discussed in the first sections.
Another issue is whether the pomeron can be treated as a `particle'
with a partonic structure, as first suggested in \cite{ingsch}, and, in
such case, what is the pomeron structure function. This question
will be addressed in the second part of this report, together with a
test of factorization.

\section{Elastic Vector Meson Production}

Elastic vector meson production in $e(\mu) p$ collisions, $e (\mu) p
\rightarrow e (\mu) V p$, is the ideal reaction to study
the interplay between the soft and the hard regime. The
electron or the muon emits a photon which fluctuates into a vector meson
and interacts with the proton via pomeron exchange.

\begin{figure}
\begin{center}
\mbox{
\psfig{figure=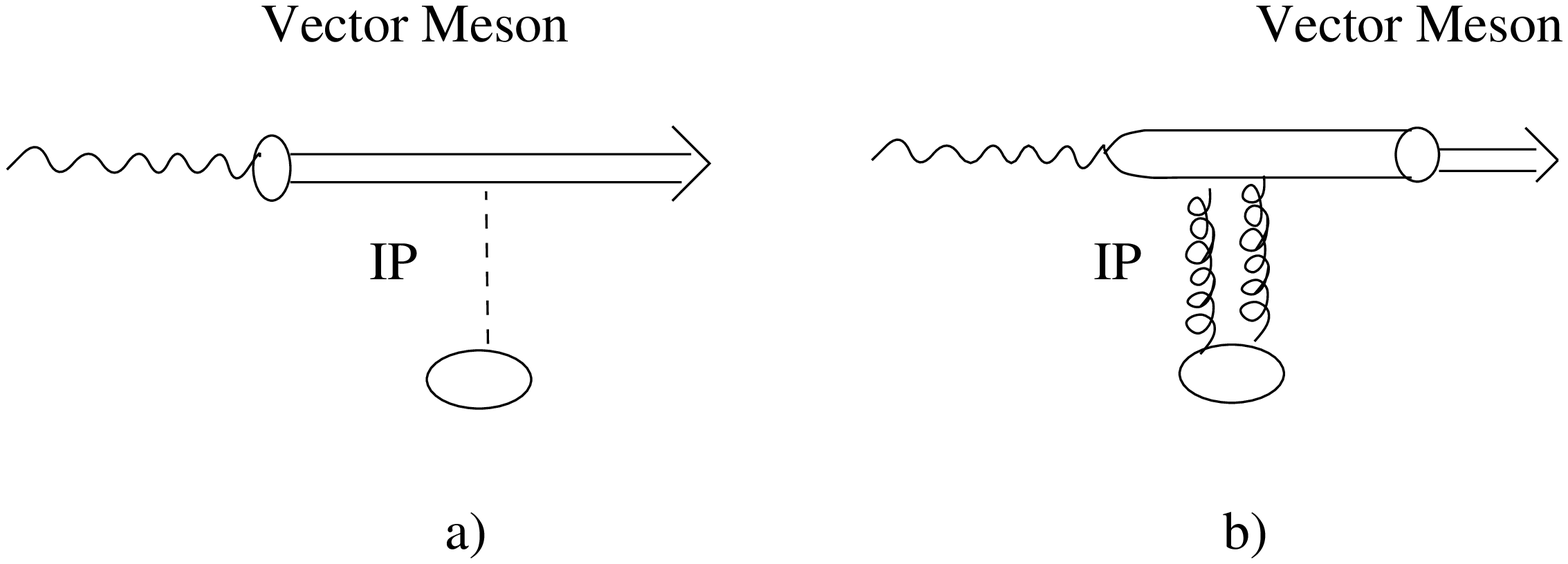,height=4cm}}
\end{center}
\caption{Elastic vector meson production in deep inelastic
scattering as seen in soft pomeron+VDM models (a) or in perturbative
QCD models (b). 
\label{fig:diagvm}}
\end{figure}

In the soft non perturbative regime \cite{softvm}, 
the process is seen as
the fluctuation of the photon into a vector meson, which 
interacts with the proton via the exchange of a soft pomeron
(see fig.~\ref{fig:diagvm}a). The dependence
of the cross section with the $\gamma p$ c.m. energy $W$ is,
assuming a pomeron intercept of 1.08 and $\alpha^\prime \simeq 0.25~\gevmtwo$,
\begin{equation}
\sigma(\gamma p \rightarrow V p) \propto
\frac {(W^2)^{2(\apom(0)-1)}} {b_0 + 2 \alpha^{\prime} \ln (W^2/W_0^2)}
                \simeq W^{0.22}.
\end{equation}

\begin{figure}
\begin{center}
\mbox{
\psfig{figure=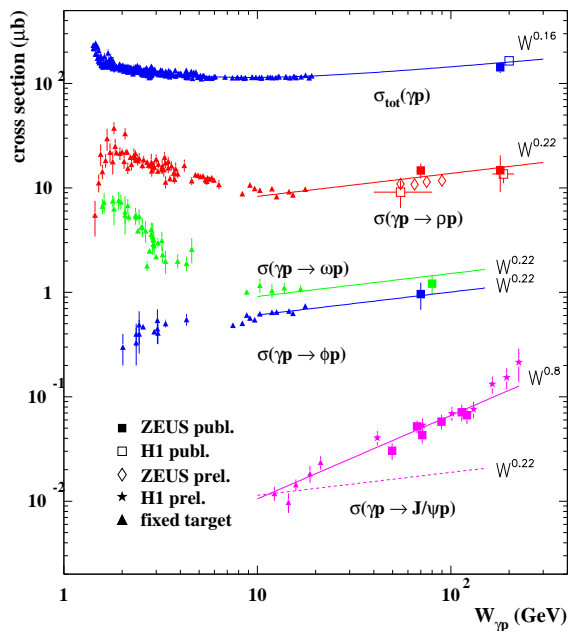,height=9cm}}
\end{center}
\caption{Cross sections for elastic vector meson production in photoproduction 
as a function of $W$.} 
\label{fig:crossvm}
\end{figure}

However, whenever a hard scale is involved in the process, which could either
be the virtuality of the photon $Q^2$, the mass of the quark in the vector 
meson or the square of the four momentum transfer at the proton vertex $t$,
the process can be calculated in perturbative QCD (see e.g.~\cite{hardvm}). 
In this framework the pomeron is seen, at leading order, 
as a two-gluon system in a colour-singlet
state, which interacts with the $q \bar q$ system of the vector
meson (see fig.~\ref{fig:diagvm}b).
The total cross section, or the longitudinal part of
the cross section in the case of virtual photons ($Q^2>0$), is 
proportional to the square of the gluon density in the proton at low $x$:
\begin{equation}
\sigma_{(L)}(\gamma^{(*)} p \rightarrow Vp) 
\propto [x G(x,Q^2)]^2 \simeq W^{4 \lambda} \simeq W^{0.8},
\end{equation}
where $xG(x) \simeq x^{-\lambda} \simeq x^{-0.2}$ 
and $W^2 \simeq Q^2/x$ at low $x$. HERA has the possibility to study
this process by varying the different scales $Q^2$, $M_V$ and $t$ (the latter is
not discussed here).

Figure~\ref{fig:crossvm} shows a compilation of cross sections for
different vector mesons in photoproduction ($Q^2\simeq 0$) as a function
of the $\gamma p$ c.m. energy $W$. The HERA data from the H1 and ZEUS
experiments extend the  region in $W$ by almost an order of magnitude
compared to fixed target experiments.
A dependence of the form $W^{0.22}$ can describe
the behaviour of the $\sigma(\gamma p \rightarrow Vp)$ cross section with
the energy for the lightest vector mesons ($\rho^0,\omega$ and $\phi$).
However the $J/\psi$ photoproduction cross section shows definitely a steeper
dependence, evident also within the HERA data
(for instance the line $W^{0.8}$ would describe the data). 

\begin{figure}
\begin{center}
\mbox{
\psfig{figure=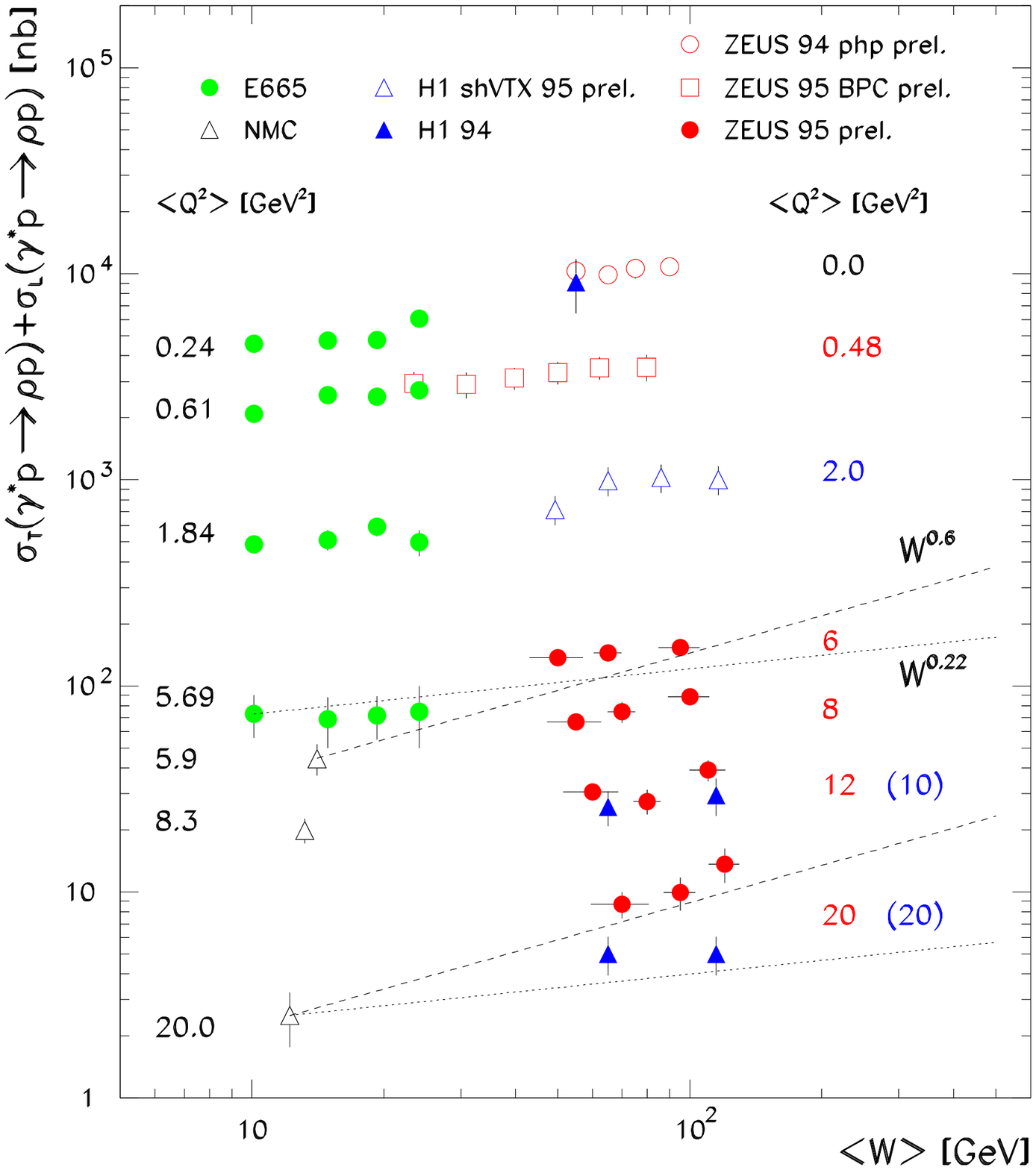,height=9cm}
\vspace{-2cm} 
\psfig{figure=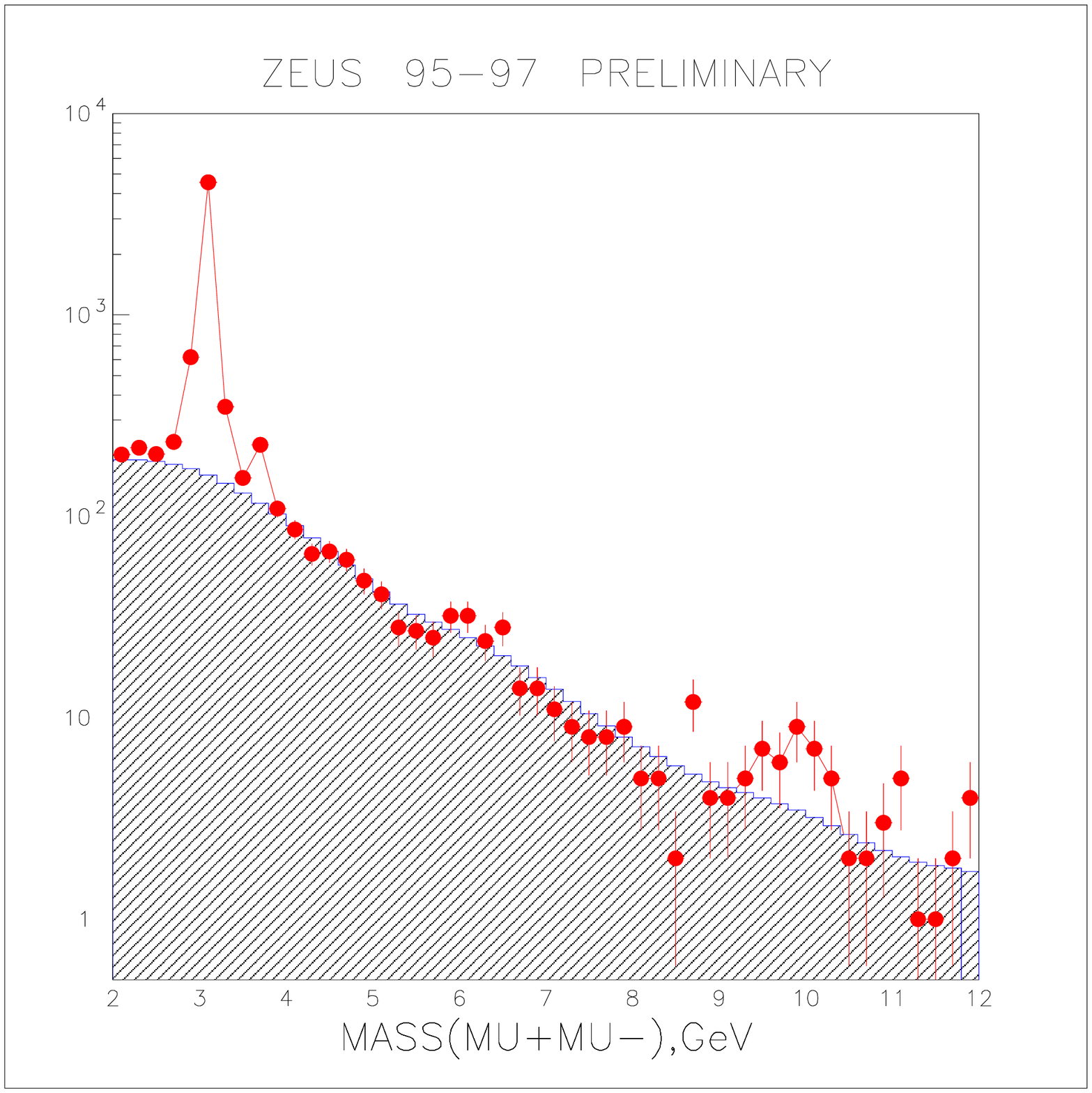,height=6cm}}
\end{center}
\caption{
On the left: cross section for $\rho^0$ production as a function of
$W$ for different $Q^2$ values. On the right:
invariant mass of $\mu^+\mu^-$ events observed in the photoproduction
95-97 ZEUS data, in the range $2-12~\gev$.
 The peaks due to the $J/\psi$, the $\psi^\prime$ and the
$\Upsilon$ resonances are visible.}
\label{fig:upsi}
\end{figure}

Figure~\ref{fig:upsi} shows a compilation of
recent results on $\rho^0$ production cross sections as a function
of the $\gamma^\ast p$ c.m. energy $W$. For low
values of $Q^2$ ($Q^2 \sleq 2~\gevtwo$) there is a nice continuation between
the recent data published by the E665 Collaboration~\cite{e665vm} 
(fixed target $\mu p$ scattering) at low $W$ 
and the preliminary HERA data at higher values of $W$:
the soft pomeron approach describes the dependence of the cross section as
a function of energy. At higher values of $Q^2$ ($Q^2 \sgeq 5~\gevtwo$) the
situation is less clear. There is some discrepancy between the E665 and
the older NMC data at $Q^2 \simeq 6~\gevtwo$~\cite{nmcvm}, which give different conclusions
when these data are extrapolated to the HERA energies and compared to the
H1 and ZEUS data. One can however look at the HERA data alone: the ZEUS
Collaboration has fitted the preliminary 95 data with the form $W^a$ for
four different $Q^2$ values, obtaining a value of $a$ which increases gradually
from $a \simeq 0.18$ at $Q^2=6~\gevtwo$ to $a \simeq 0.77$ at 
$Q^2\simeq 20~\gevtwo$, however with large errors.

\begin{figure}
\begin{center}
\mbox{
\psfig{figure=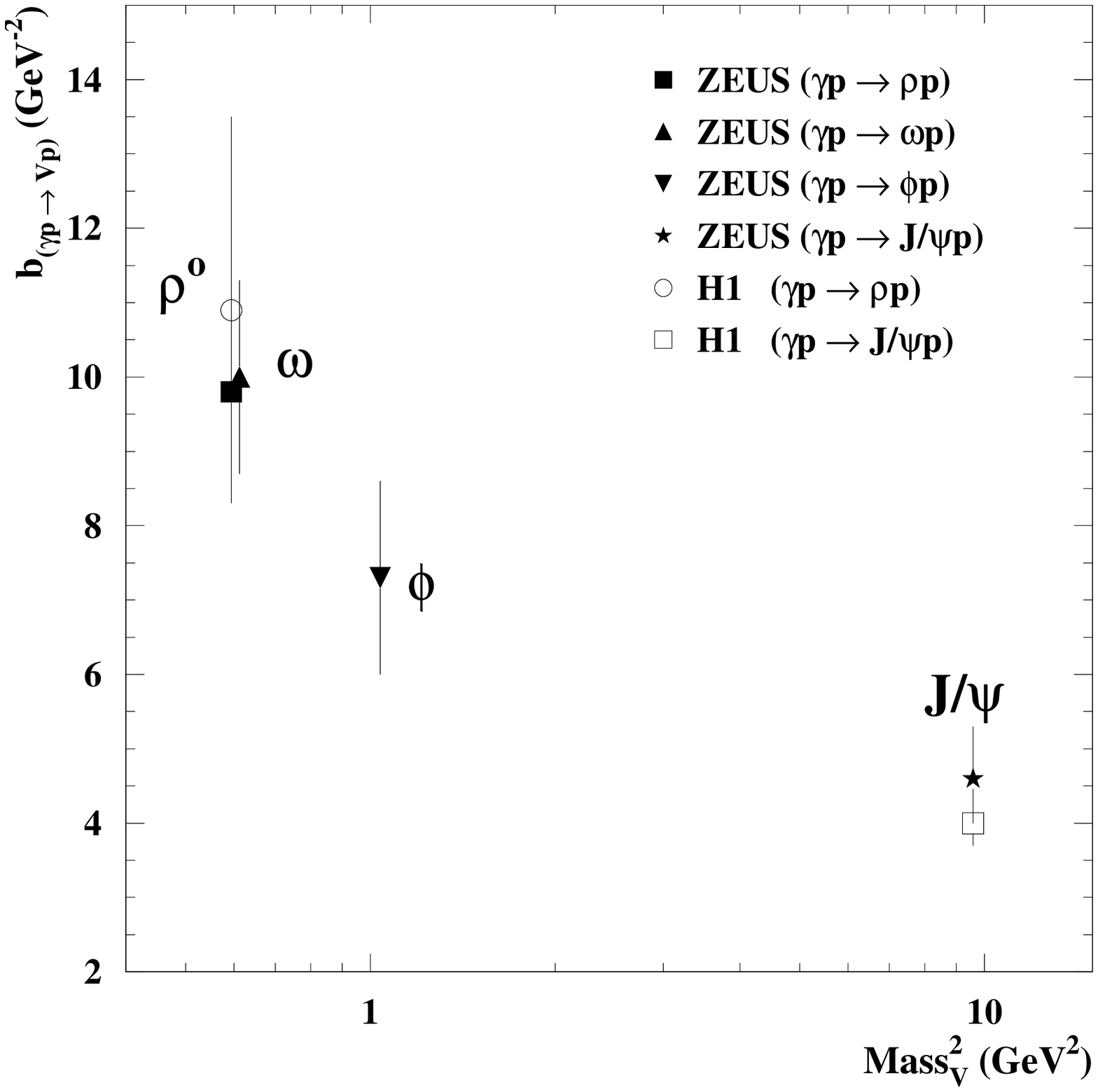,height=7cm}
\psfig{figure=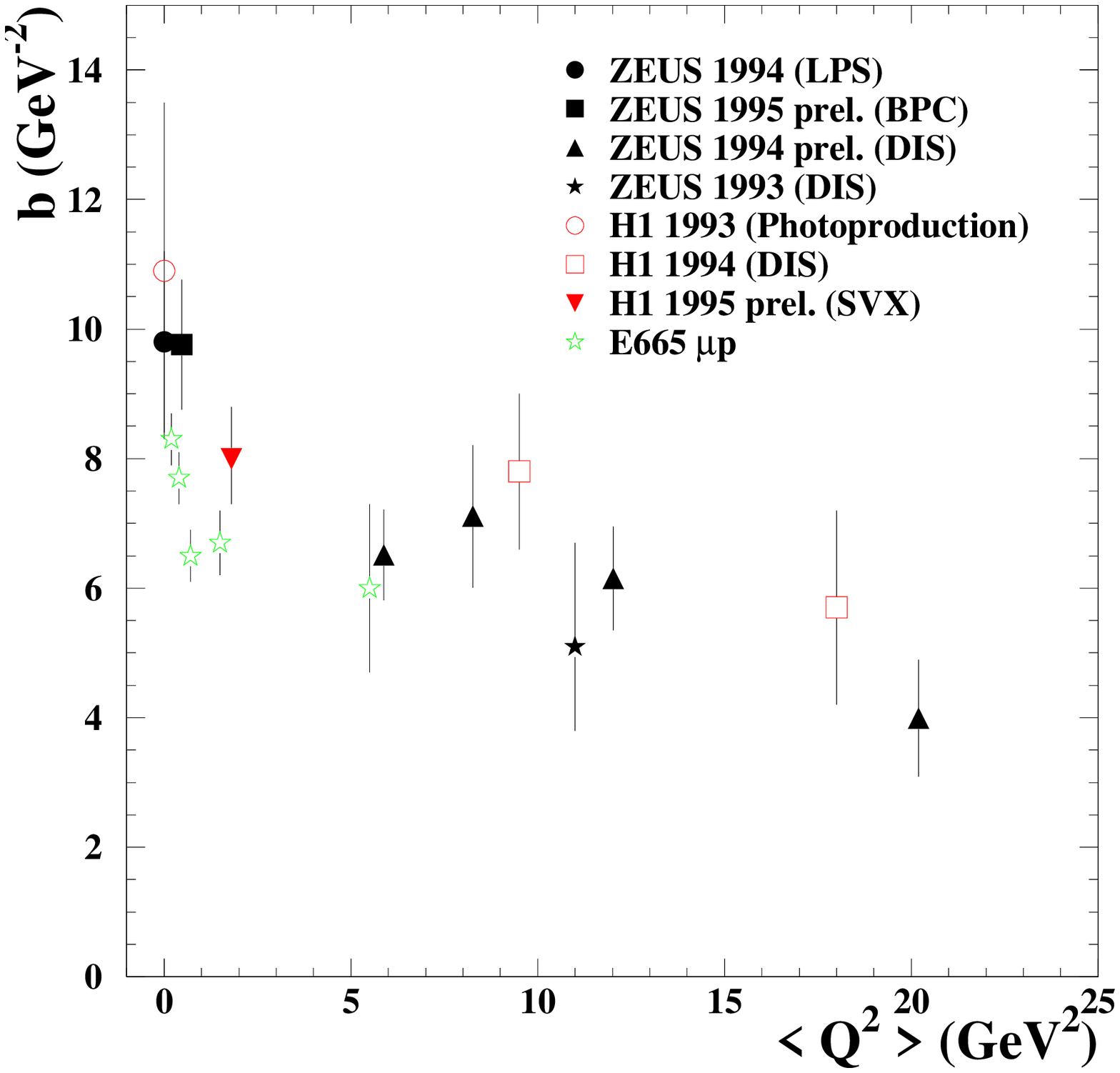,height=7cm}}
\end{center}
\caption{
Dependence of the slope parameter $b$ as a function of the mass squared
of the vector meson in photoproduction (left) and 
as a function of $Q^2$ for the $\rho^0$ (right).}
\label{fig:slopevm}
\end{figure}

In summary the mass of the charm in the $J/\psi$ photoproduction 
may give the hard
scale, causing the steep rise of the cross section with $W$, while for
the production of the light vector meson $\rho^0$ there is an indication
that increasing $Q^2$ perturbative QCD starts to play a role, even
if there are still experimental uncertainties. In principle this process
is a very important reaction to determine the gluon density at low $x$,
as the cross section has a quadratic dependence on $xG(x,Q^2)$, but there are
still theoretical uncertainties in this determination.
It will be interesting to study further this process with more
statistics and with other vector mesons. The ZEUS Collaboration has 
observed in the 1995-1997 photoproduction data 
of the order of 20 $\Upsilon$ events in the
$\mu^+\mu^-$ channel (fig.~\ref{fig:upsi}). The cross
section, assuming that most events originate from the $\Upsilon(1S)$ state,
is:
\begin{equation}
\sigma(\gamma p \rightarrow \Upsilon N, M_N \sleq 4~GeV) =  
                0.9 \pm 0.3 \pm {0.3}~nb,  
\end{equation}
in the kinematic range $80 <W < 280~\gev$ and $Q^2<4~\gevtwo$,     
where $N$ indicates the proton or the low mass state in which the proton
dissociates. For $M_N \sleq 4~\gevtwo$ this proton dissociation background 
contributes for 
$\sleq 30\%$ to the elastic $\Upsilon$ production. The cross section is of
the order of $1\%$ of the $J/\psi$ photoproduction cross section, as
predicted in some perturbative QCD models~\cite{frankheavy}.

Another interesting variable to investigate the role played by 
perturbative QCD is the slope parameter $b$ in the exponential
$t$ dependence of the cross section 
($d \sigma/dt \propto e^{-b|t|}$ at low $|t|$). 
This is shown in fig.~\ref{fig:slopevm} as a function
of the mass of the vector meson in photoproduction and 
as a function of $Q^2$ for the $\rho^0$.
The slope parameter $b$, which is related to the transverse size of the interaction
($b \propto R_p^2 + R_V^2$, where the radius of the proton $R_p^2$ corresponds
to $\simeq 4~\gevmtwo$),
is seen to decrease with the mass or with $Q^2$. This corresponds to the
pQCD picture in which, at low $x$, the photon fluctuates into the $q \bar q$
system well before the target; the proton then interacts via the two-gluon
exchange with the $q \bar q$ system, whose transverse size decreases with
increasing $Q^2$ or the mass of the vector meson. The process then becomes
a short distance process and  therefore calculable in pQCD.

\section{Single Photon Dissociation at HERA}

\begin{figure}
\begin{center}
\mbox{
\psfig{figure=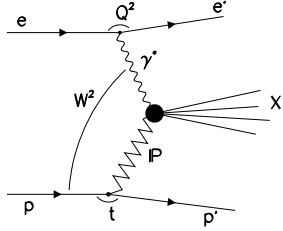,height=4cm}}
\end{center}
\caption{
Feynman diagram for single dissociation at HERA.
\label{fig:diagsd}}
\end{figure}

The single photon dissociation process at HERA, $ep \rightarrow e Xp$,
is an important tool to study the nature of the pomeron and
 its possible partonic structure. The diagram of the process is drawn 
in fig.~\ref{fig:diagsd}: the electron emits a 
quasi-real photon ($Q^2 \simeq 0$) or a virtual photon (for $Q^2>0$)
which interacts with the proton via exchange of a pomeron 
and dissociates into a system $X$. The square of the four-momentum transfer $t$
at the proton vertex is typically small ($|t| \sleq 1~\gevtwo$) and the
scattered proton carries almost all the initial beam momentum. As the pomeron
is a colour-singlet state, a rapidity gap in particle flow is expected
between the scattered proton direction and the system $X$. These
signatures are used to select inclusive diffractive events. 

H1 selects diffractive events by requiring a large rapidity gap between
the system $X$ and the system $N$ going into the forward proton direction,
where $N$ can either be the scattered proton or the state into which the
proton dissociates. The forward pseudo-rapidity coverage of H1 is 
$3.4< \eta < 7.5$~\footnote{The pseudo-rapidity $\eta$
is defined as $-\ln (\tan (\theta/2))$, where $\theta$ is the polar angle
measured with respect to the proton direction.},
constraining the mass of the system $N$ to be less than $1.6~\gev$. The cross
section that H1 quotes contains thus a $5\%$ contamination due to
double dissociation. ZEUS uses instead two different methods for
the selection of single $\gamma^{(\ast)}$ dissociation events. The first method
exploits the different behaviour of the $M_X$ distribution for non diffractive
and diffractive events. The $\ln M_X^2$ distribution can be written
as the sum of two components: 
at high $M_X$, the $\ln M_X^2$ distribution exhibits an exponential
fall-off for normal DIS events; 
at low $M_X$ it is approximately flat as expected for
diffractive events. The second method is based instead on the measurement
of the scattered proton in the leading proton spectrometer (LPS),
requiring a proton with a fraction of the  incoming beam energy $x_L$ close to 1.
This latter method has the advantage to allow 
a measurement of $t$, but the acceptance is small.

\subsection{Photoproduction}

Following Regge phenomenology\cite{mueller}, the double differential cross
section for  the process $\gamma p \ra Xp$ can be written as:
\begin{eqnarray}
\frac{d \sigma}{dt dM_X^2} & = &
(\frac{1}{M_X^2})^{\apom(0)} (W^2)^{2 \apom(0) -2} e^{-b|t|} 
\label{eq:apomphp}
\\
b & = & b_0 + 2 \cdot \alpha^\prime \ln (W^2/M_X^2).
\label{eq:tphp}
\end{eqnarray}

The dependence of the diffractive $\gamma p$ cross section on $t$ has been
measured by ZEUS in the kinematic range $176<W<225~\gev$,
$4<M_X<32~\gev$, $0.07<|t|<0.4~\gevtwo$ and $0.97<x_L<1.02$.
The cross section exhibits an exponential shape, and a fit
with the form $e^{-b|t|}$ yields the value for the slope parameter
$b=7.3 \pm 0.9 \pm 1.0~\gevmtwo$ (preliminary).
The value can be compared to earlier results obtained in $\gamma p$
interactions by the E612 Collaboration~\cite{chapin}, at lower
values of $W$ ($<W>=14~\gev$) and in a different $t$
range ($0.02<|t|<0.1~\gevtwo$). The formula~(\ref{eq:tphp}) with
$\alpha^\prime=0.25~\gevmtwo$ describes the dependence of $b$ on 
the ratio $W^2/M_X^2$; however the experimental errors are still large.

The dependence of the differential cross section on $M_X^2$ gives
instead a measurement of the pomeron intercept at $Q^2 \simeq 0$.
The measurement has been performed both by H1 and ZEUS at 
$W\simeq 200~\gev$, obtaining
$\apom(0)=1.07 \pm 0.02(stat.) \pm 0.02 (syst.)\pm 0.04(model)$ (H1)  
\cite{h1php}
and $\apom(0)=1.12 \pm 0.04(stat) \pm 0.08 (syst) $ in
the range $ 8< M_X < 24~\gev$ (ZEUS)~\cite{zeusphp}. 
Both values are compatible with the soft pomeron intercept.

\subsection{Deep Inelastic Scattering}

As already mentioned, the measurement of inclusive single diffractive
dissociation in DIS can give information on the nature of the pomeron
and, assuming that it can be treated as a particle, the virtual photon
can probe its partonic
structure. The kinematics of the process (see fig.~\ref{fig:diagsd})
is described, in addition to the usual DIS variables $Q^2,x$ 
and $y=Q^2/xs$, by the two variables:
\begin{equation}
\xpom \simeq \frac{M_X^2+Q^2}{W^2+Q^2} \simeq 1-x_L~,~~
\beta \simeq \frac {Q^2}{Q^2+M_X^2}
\end{equation}
as well as by $t$. For pure pomeron exchange, $\xpom$ is the fraction
of the proton momentum carried by the pomeron and $\beta$ can be interpreted
as the fraction of pomeron momentum carried by the struck quark (the 
analogue of $x$ Bjorken for the proton).
The cross section for this process can be written, in analogy to normal
DIS events, in terms of a
diffractive structure function $\ftwodfour$:
\begin{equation}
\frac{d^4 \sigma^D_{(ep \rightarrow e^\prime Xp^\prime)}} {d \beta d Q^2 d \xpom dt}=
\frac{4 \pi \alpha^2}{\beta Q^2} (1-y+ \frac{y}{2})
{F_2^{D(4)}}(\beta,Q^2,\xpom,t) +\delta_L + \delta_Z,
\end{equation}
where $\delta_L$ and $\delta_Z$ are the contributions due to the longitudinal
structure function and to the $Z$ exchange, respectively. These will be
neglected in the kinematic range of the measurements described here.
In factorizable models where the pomeron is treated like a particle
(see e.g.~\cite{ingsch}), the diffractive structure function can be
written as the product of two terms:
\begin{equation}
{F_2^{D(4)}} = {f^{\pom}(\xpom,t) F_2^{\pom}(\beta,Q^2)},
\label{eq:fact}
\end{equation}
 where the the flux $f^{\pom}$ depends only on $\xpom$ and $t$ and $ F_2^{\pom}$,
 which can be interpreted as a pomeron structure function,
depends only on $\beta$ and $Q^2$.
In most of the measurements the scattered proton is usually not detected
and $t$ cannot be measured, therefore the structure function
$\ftwod$ is defined, obtained by integrating $\ftwodfour$ over $t$. 
Equation~(\ref{eq:fact})
becomes then
\begin{equation}
{F_2^{D(3)}} \propto \frac{1}{\xpom^a} F_2^{\pom}(\beta,Q^2),
\label{eq:xpom}
\end{equation}
where the $\xpom$ exponent can be related to the pomeron intercept
$\overline {\apom}$, integrated over the $t$ range, with the relation
$a= 2\overline{\apom}-1$. If factorization holds, $a$ is independent of
$\beta$ and $Q^2$.

\begin{figure}
\begin{center}
\mbox{
\psfig{figure=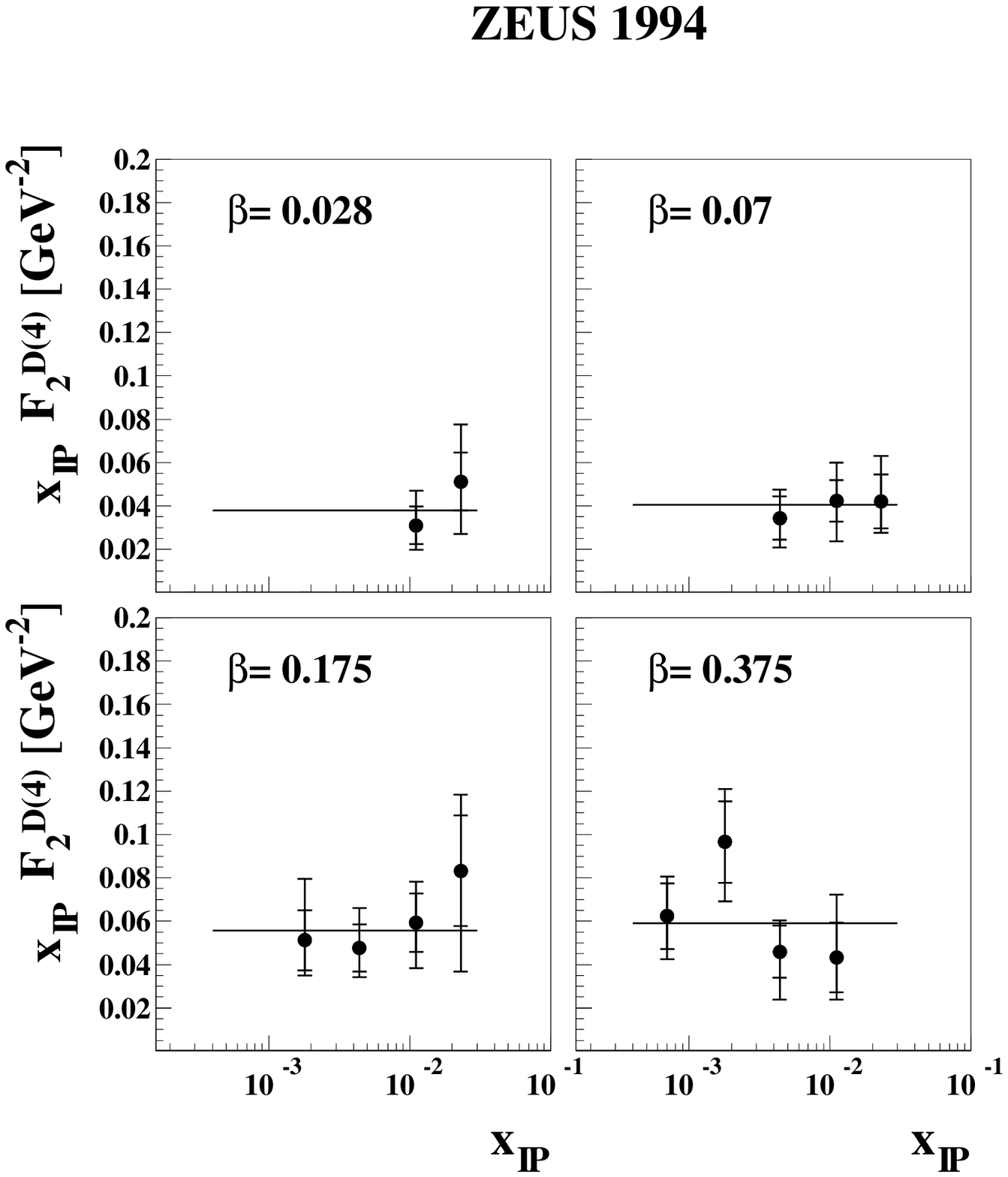,height=7cm}}
\end{center}
\vspace{-.5cm}
\caption{Preliminary ZEUS results on the structure function
$\ftwodfour$.
\label{fig:zeusftwod4}}
\end{figure}

\begin{figure}
\begin{center}
\mbox{
\psfig{figure=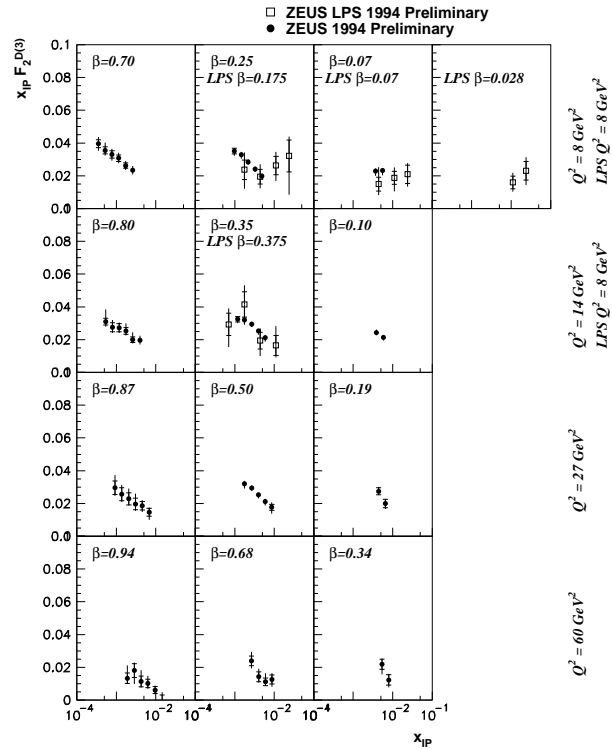,height=11cm}}
\end{center}
\vspace{-1cm}
\caption{Preliminary ZEUS results on the structure function
$\ftwod$ obtained with the $M_X$ method (black dots) and with the LPS
tag (open squares). 
% The data obtained with the $M_X$ method
% have been scaled down by $25\%$ in this comparison, to take into account the
% contribution due to the proton dissociation background. This type of
% background is instead negligible in the LPS data. 
\label{fig:zeusftwod3}}
\end{figure}

ZEUS has measured the structure function $\ftwodfour$, using the 1994
data in which the scattered proton was measured in the LPS, in the
kinematic range $x_L>0.97$,$5<Q^2<20~\gevtwo$,$0.015<\beta<0.5$ and
$0.073<|t|<0.4~\gevtwo$. The result is shown in fig.~\ref{fig:zeusftwod4}
where the function $\xpom \ftwodfour$ is plotted as a function of $\xpom$
in four $\beta$ bins. Figure~\ref{fig:zeusftwod3} shows
the structure function $\ftwod$ plotted as $\xpom \ftwod$ as a function of 
$\xpom$, obtained by integrating $\ftwodfour$ over $t$.
Together with the LPS data, the points
obtained with the $M_X$ method are also shown.
The LPS data extend the measurements to higher values
of $\xpom$ and  lower values of $\beta$, as higher values of $M_X$ can be
reached with this method. The
two measurements agree in the region of overlap both in shape and 
in normalization. The $\xpom$ dependence seems to change with the
$\xpom$ range, having a negative slope for the function $\xpom \ftwod$
at low $\xpom$ values, and becoming flat or even positive at higher $\xpom$.

\begin{figure}
\begin{center}
\mbox{
\psfig{figure=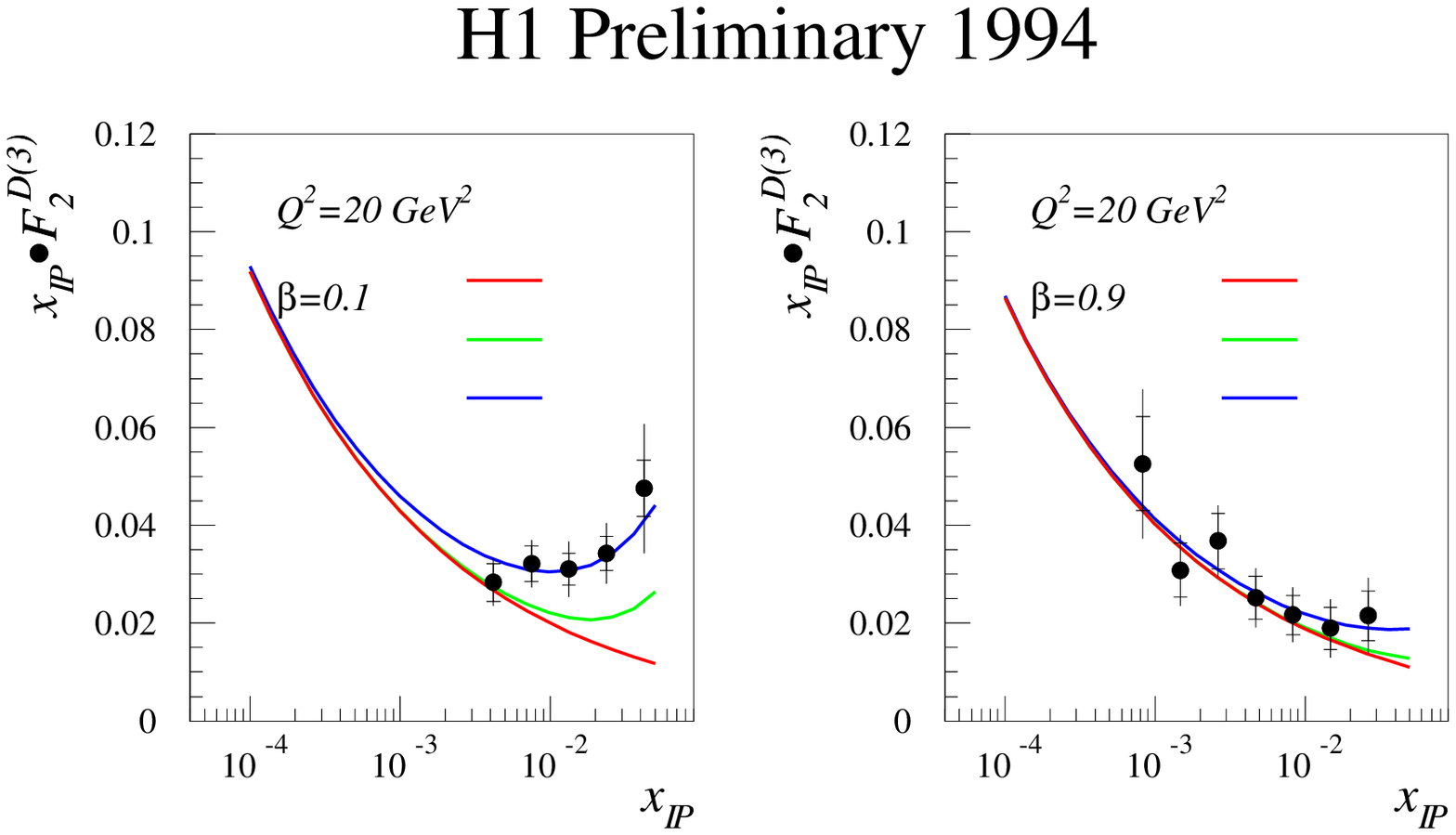,height=6cm}}
\end{center}
\vspace{-1cm}
\caption{Preliminary H1 results on the structure function
$\ftwod$ in two $\beta$ bins (0.1 and 0.9) at $Q^2=20~\gevtwo$. 
The data (black dots) are fitted with the sum (upper line) of a pomeron
contribution (lower line) and a reggeon contribution with interference
(middle line).
\label{fig:h1ftwod3}}
\end{figure}

This dependence of the $\xpom$ slope on the kinematic range was first
observed by the H1 Collaboration in their preliminary 1994 data~\cite{h1warsaw},
 where the structure function $\ftwod$ was measured with a fine binning in
a wide kinematic range ($2.5<Q^2<65~\gevtwo$,$0.01<\beta<0.9$,
$0.0001< \xpom< 0.05$). 
The $\xpom$ dependence was found to change with $\beta$,
as illustrated in two bins in fig.~\ref{fig:h1ftwod3}.
One of the possible explanations is factorization breaking effects,
predicted by some pQCD models (see e.g.~\cite{factbreak}). 
The other possibility is the contribution of additional reggeon exchanges 
at high values of $\xpom$: as the reggeon has an intercept 
$\alphareg(0) \simeq 0.5$, its effect is to lower the effective exponent $a$ 
in the $1/\xpom$ dependence in equation~(\ref{eq:xpom}). H1 has made a fit
to all bins, parametrizing $\ftwod$ as a sum of a pomeron
and a reggeon contribution:
\begin{equation}
{\ftwod} = {F_2^{\pom}(\beta,Q^2)} \cdot 1/\xpom^{(2 \apom -1)}
              +C_{\reggeon} {F_2^{\reggeon}}(\beta,Q^2) \cdot 1/\xpom^{(2 \alphareg-1)},
\end{equation}
where for the reggeon the pion structure function as parametrized in
\cite{grvpion} has been used.
An example of fit, where the $f$ interference is included, is shown
in fig.~\ref{fig:h1ftwod3}. The data are described, 
in all $\beta$ and $Q^2$ bins, by
a pomeron intercept $\apom(0)=1.18 \pm 0.02(stat) \pm 0.04 (syst.)$ and
a meson intercept $\alphareg(0)=0.6 \pm 0.1 \pm 0.3 $.  
The introduction of these  additional subleading trajectories restores factorization.

\begin{figure}
\begin{center}
\mbox{
\psfig{figure=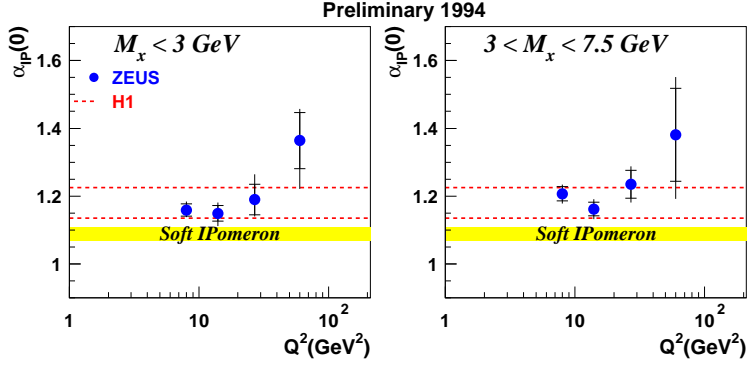,height=5cm}}
\end{center}
\caption{Preliminary H1 and ZEUS results on the 
pomeron intercept $\apom(0)$ in two $M_X$ bins and as a function of $Q^2$.
The dots are the ZEUS data, the dashed line represent the H1 value 
(with
the statistical and systematic errors added in quadrature) obtained
from the pomeron plus reggeon fit. 
The shaded band represents a soft pomeron
possible value ranging from 1.07 to 1.11.
\label{fig:apom}}
\end{figure}

\begin{figure}
\begin{center}
\mbox{
\psfig{figure=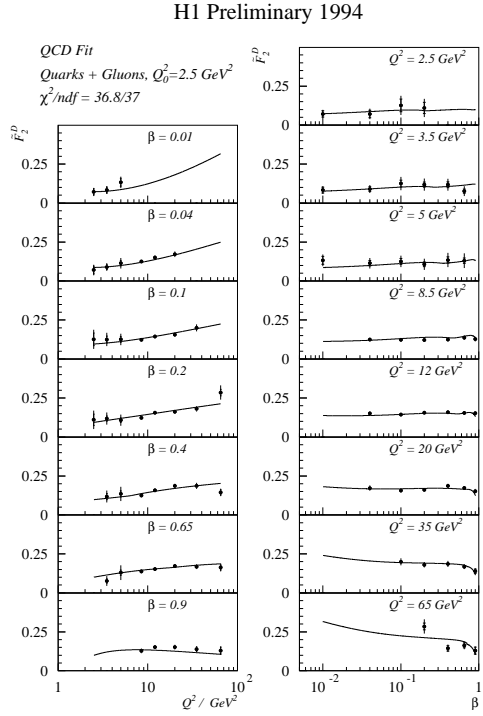,height=10cm}}
\end{center}
\caption{
The structure function $F_2^D$ from the H1 1994 preliminary data.
%as a function of $Q^2$ in different $\beta$ bins, and as a function of
%$\beta$ in different $Q^2$ bins. 
The solid line indicates the result of
a QCD fit including quarks and gluons.
\label{fig:h1ftwod2}}
\end{figure}

The ZEUS Collaboration has instead fit the data obtained with
the $M_X$ method at low values of
$\xpom$ ($\xpom \sleq 0.01$), where reggeon exchange contributions are
expected to be small, in two $M_X$ bins and for the four different $Q^2$
values (see fig.~\ref{fig:zeusftwod3}). The result on $\apom(0)$ is shown
in fig.~\ref{fig:apom}, where the dots indicate the ZEUS results, while
the dashed line indicates the H1 result. The two experiments agree;
the ZEUS data may indicate a variation of the intercept
with $Q^2$, however the errors are still large to draw any conclusion.
The values are above the soft pomeron intercept, which is indicated in the
figure with a band in the range $1.07-1.11$. A value of $\apom(0) \simeq 1.2$ 
as obtained in these DIS data is
higher than the value obtained at $Q^2\simeq 0$; it also
leads to the same steep $W$ dependence ($\simeq W^{0.8}$)
as measured in the elastic $J/\Psi$ photoproduction data. These facts suggest
that pQCD may play a role when the $Q^2$ is large.

\begin{figure}
\begin{center}
\mbox{
\vspace{-8cm}
\psfig{figure=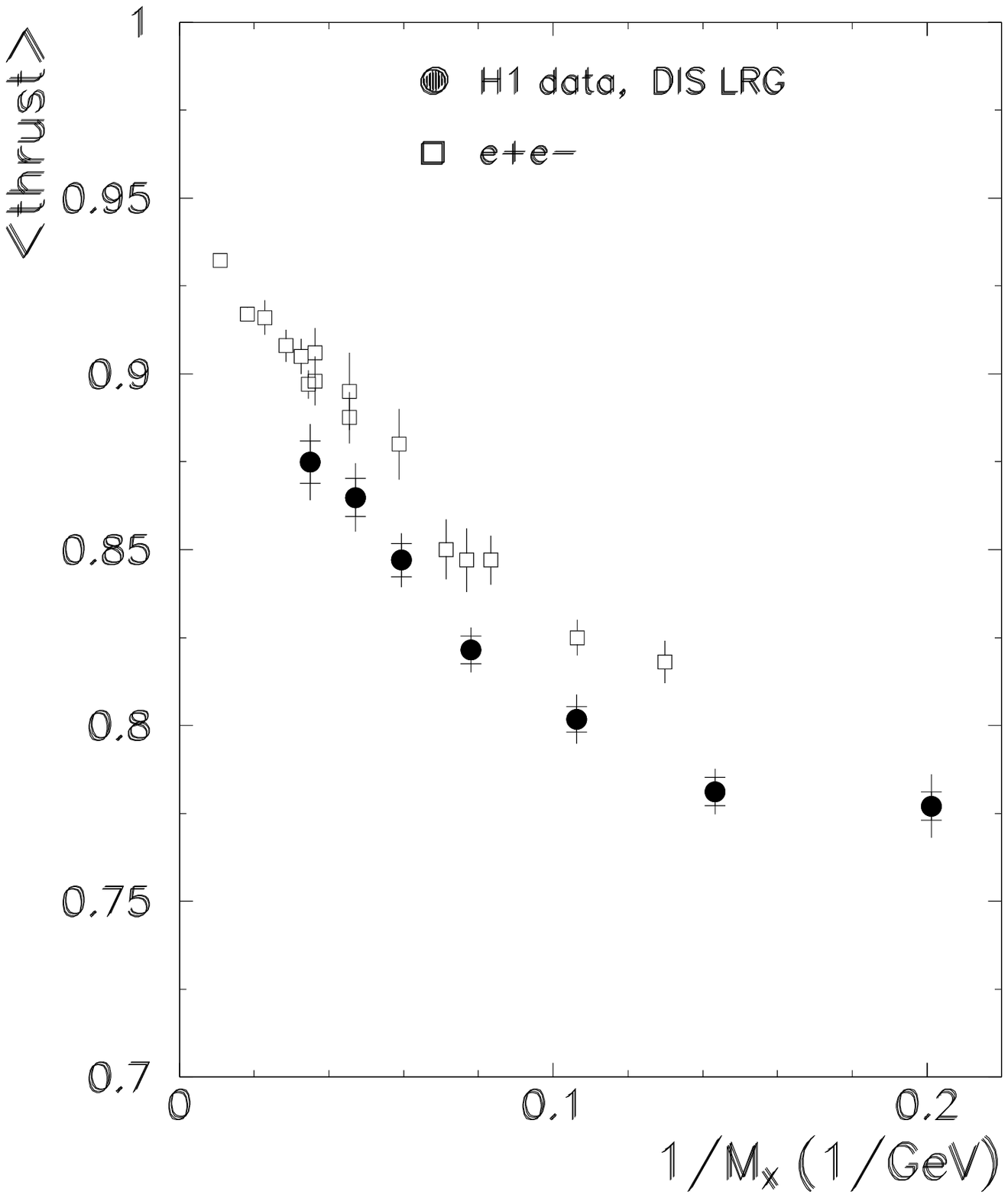,height=9cm}
\vspace{.5cm}
\hspace{-0.5cm}
\psfig{figure=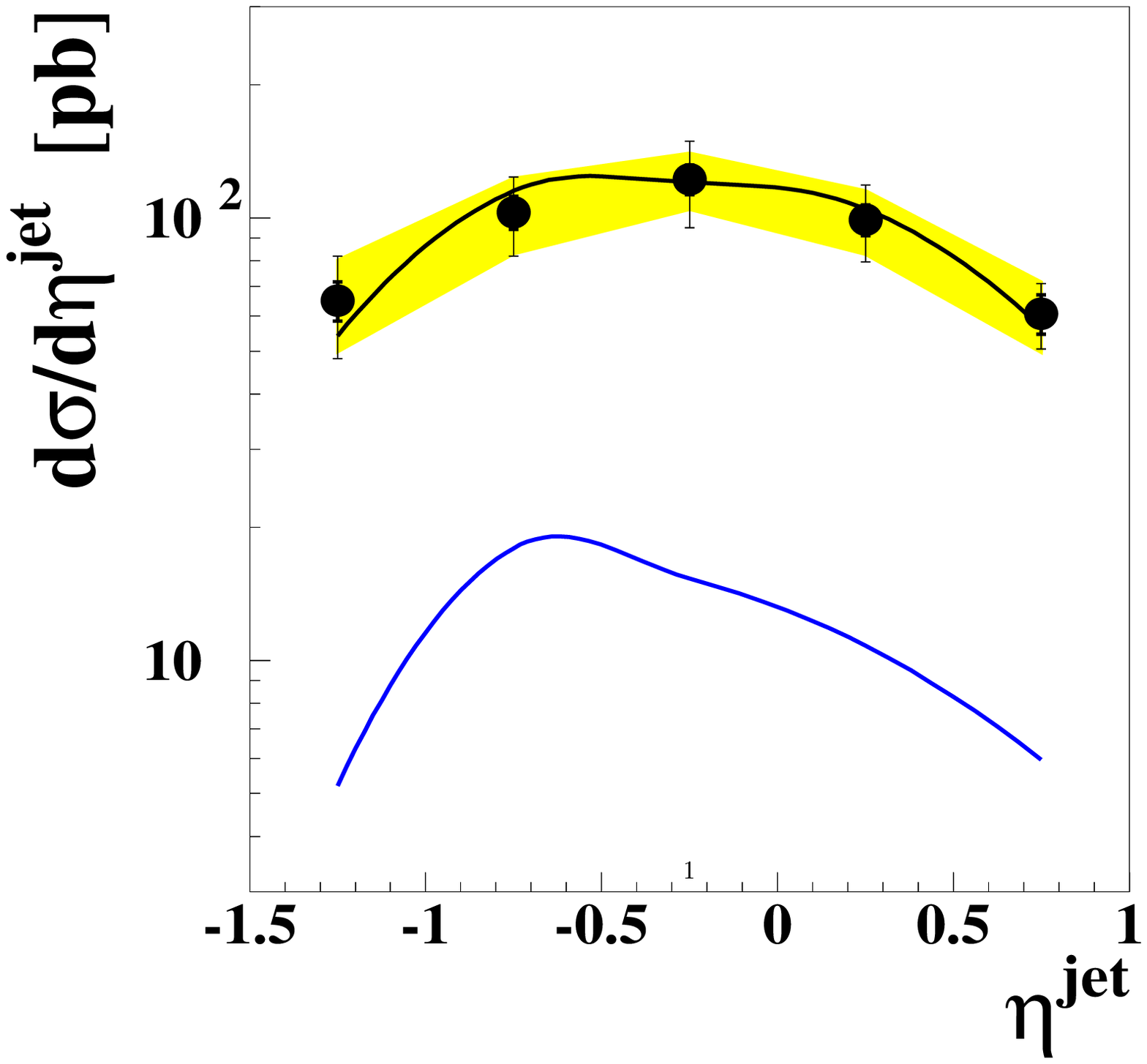,height=10cm}
}
\end{center}
\vspace{-.5cm}
\caption{
On the left: the mean value of the thrust  in the $\gamma^\ast \pom$ system
as a function of $1/M_X$,
obtained from the H1 1994 data (dots), compared to $e^+ e^-$ data at 
similar c.m. energies.
On the right: the differential cross section for
dijet events in diffractive $\gamma p$ interactions as a function of the jet pseudorapidity,
 as measured by ZEUS in the preliminary 1994 data. 
The points are the data, the dashed band is the
systematic error due to the calorimeter energy scale, the upper solid
line is the prediction for a hard gluon structure function ($\beta(1-\beta)$),
the lower solid line is the prediction corresponding to a hard quark structure function
($\beta(1-\beta)+c (1-\beta)^2$) for the pomeron.}
\label{fig:evshapes}
\end{figure}

\begin{figure}
\begin{center}
\mbox{
\psfig{figure=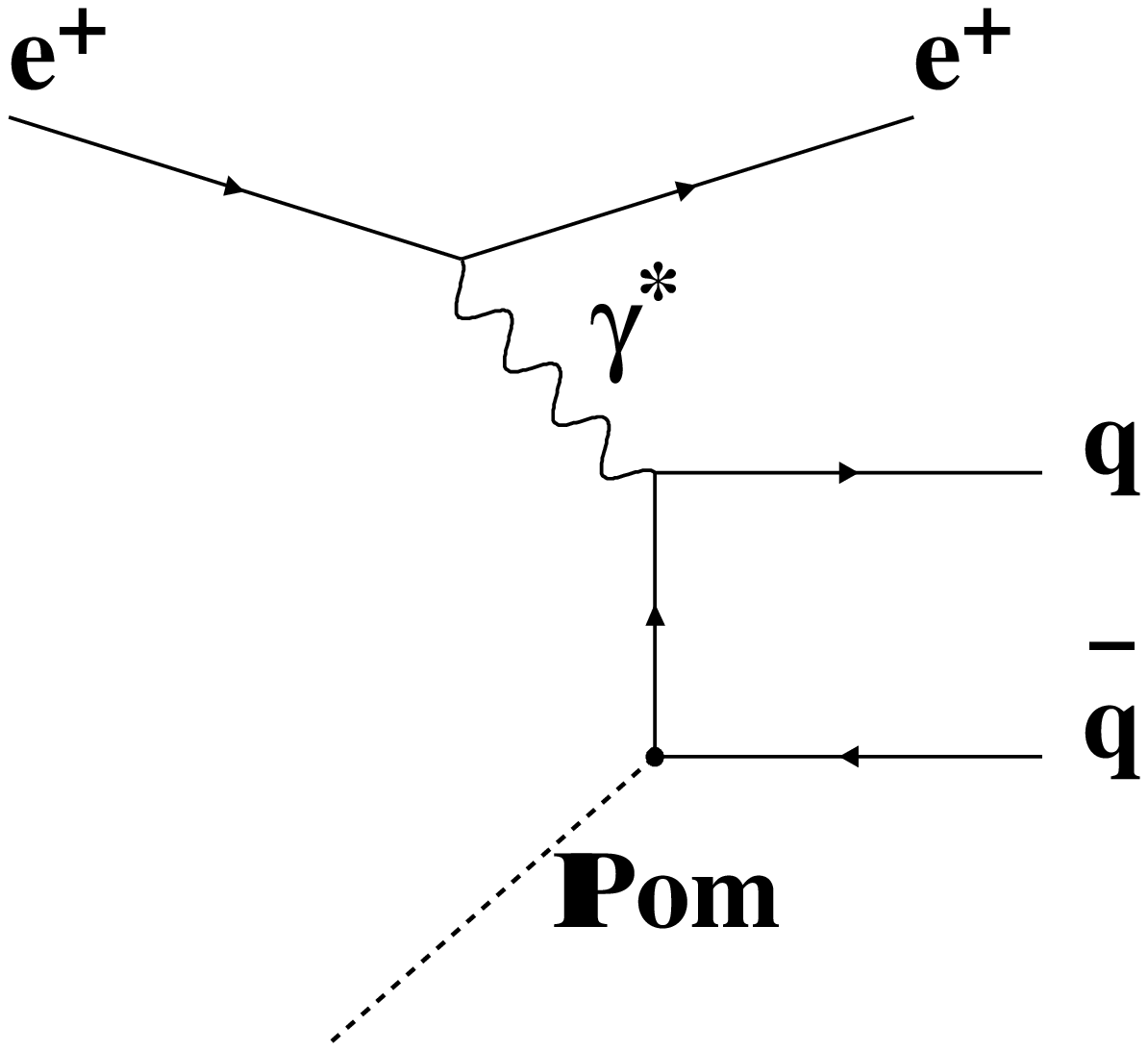,height=4cm}
\psfig{figure=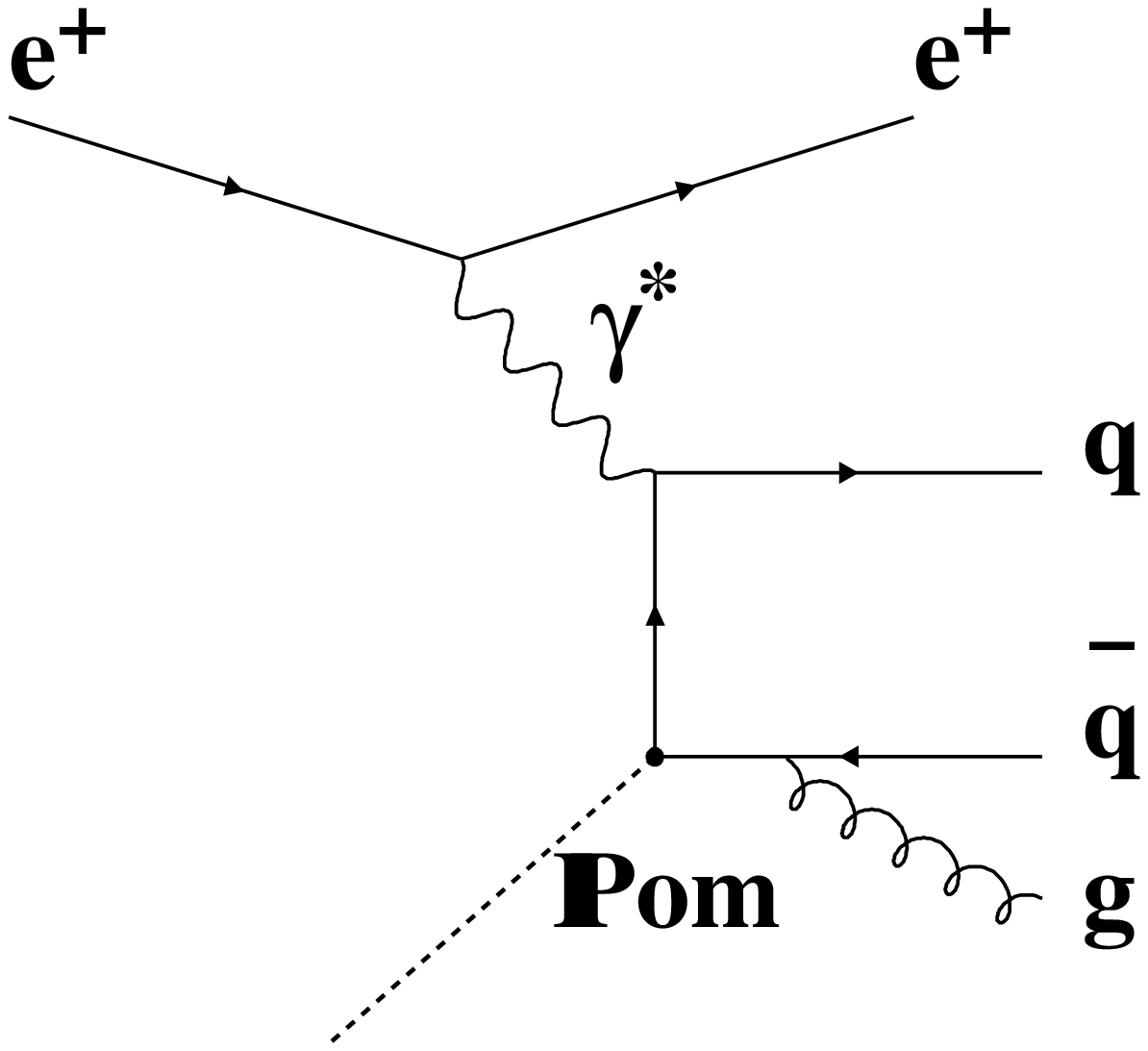,height=4cm}
\psfig{figure=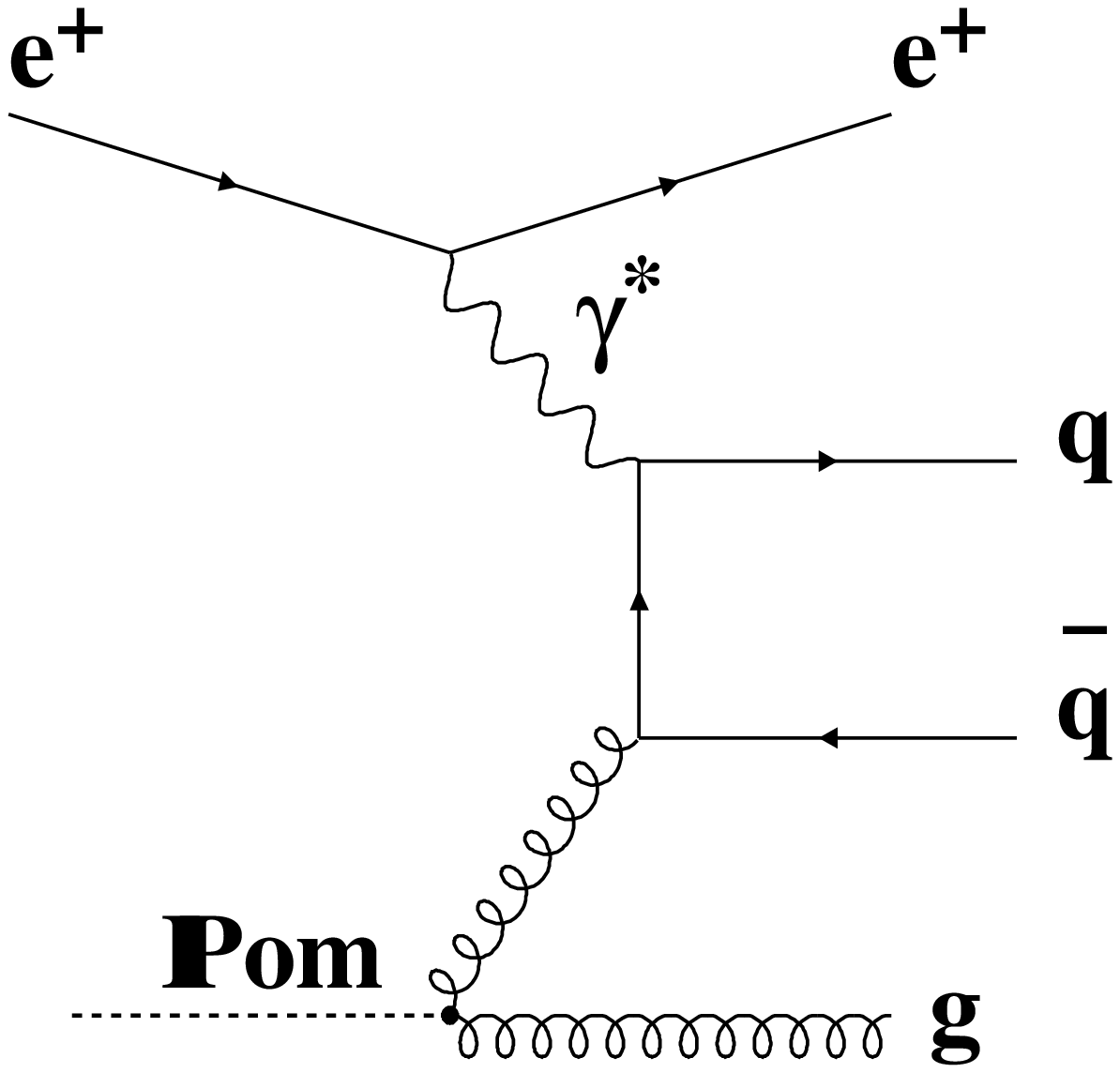,height=4cm}
}
\end{center}
\vspace{-.3cm}
\caption{
Feynman diagrams for diffractive DIS single dissociation, including
the lowest order, and the two first leading order diagrams, QCD Compton
and BGF.
\label{fig:diaglo}}
\end{figure}

H1 has also studied the pomeron structure function as a function of
$\beta$ and $Q^2$. Integrating $\ftwod$ over the measured $\xpom$ range, they obtained:
\begin{equation}
{ {\tilde{F}}_2^D(\beta,Q^2) = \int^{\xpom=0.05}_{\xpom=0.0003} \ftwod(\beta,Q^2,\xpom)
     d \xpom}, 
\end{equation}
where, assuming factorization,
 the structure function $\tilde{F}_2^D$ is proportional to the pomeron structure function.
The latter can be written as a sum over the quark densities in the pomeron,
$F_2^{\pom}(\beta,Q^2)=\beta \sum_q e^2_q q(\beta,Q^2)$.
The result is shown in fig.~\ref{fig:h1ftwod2} as a function of $\beta$ and $Q^2$,
where the black dots represent the H1 data. One observes scaling violations
in the $Q^2$ dependence and a relatively flat dependence versus $\beta$. The
scaling violations have a positive slope even at very high values of $\beta$
(the ``valence quark'' region),
as expected if the gluon component dominates.
H1 has performed a QCD fit to the data, assuming initial parton densities
of the form $q_i,g=A_i x_{i/{\pom}} (1- x_{i/{\pom}})^C_i$ at a starting value of
$Q_0^2=2.5~\gevtwo$, and evolving them at higher $Q^2$ with the DGLAP equations.
A first fit, assuming
that only quark densities contribute to the pomeron structure function, gave
a very poor description of the data. A second fit, which includes also the gluon,
is shown as the solid line in fig.~\ref{fig:h1ftwod2} and describes the data
well. The gluon component thus obtained is found to be hard, 
% that is almost all
% the pomeron momentum is carried by the parton, 
and it varies from 
approximately $90\%$ at low $Q^2$ to $80\%$ at higher $Q^2$ values.
Note that this dominant gluon component is in agreement with models where rapidity
gap events at HERA are described in terms of the boson gluon fusion (BGF)
process, in which the final state $q \bar q$ system evolves in a colour-singlet
state, 
creating a rapidity gap with respect to the proton
remnant \cite{sci}.

\subsection{Final states}

The measurement of the structure function $\ftwod$ gives information on
the gluon component in the pomeron through the scaling violations. A
complementary information can be obtained by studying specific hadronic final states.
The H1 and ZEUS experiments have studied various aspects of the final states,
here I will mention only two examples. 

The H1 Collaboration has measured the thrust distribution in diffractive
DIS events in the $\gamma^\ast \pom$ system. The mean value of the thrust
is shown as a function of $1/M_X$ in fig.~\ref{fig:evshapes}. The mean thrust
is seen to increase as $M_X$ increases, as expected if the events become
more collimated. However, when the H1 values are compared to the mean thrust
measured in $e^+ e^-$ collisions at similar c.m. energies ($\sqrt{s_{e^+ e^-}}
\simeq M_X$)~\cite{pluto}, 
the diffractive events seem to be broader, as the mean thrust is always
lower. This can be understood by considering the Feynman diagrams in leading
order (fig.~\ref{fig:diaglo}): while for $e^+ e^-$ processes only
the analogue of the first two diagrams is present, the diffractive single dissociation
process contains also the last diagram from the gluon component in the
pomeron.

Dijet production in photoproduction is also sensitive to this last diagram and 
therefore to the gluon content in the
pomeron.
The ZEUS experiment has measured the differential cross section
$d \sigma/ d \eta^{\mathrm jet}$ for dijet events at $Q^2 \sleq 4~\gevtwo$ and
$E_T^{\mathrm jet}> 6~\gev$, as a function of the jet pseudorapidity
$\eta^{\mathrm jet}$. The data are shown  
in fig.~\ref{fig:evshapes}, where they are compared to two different
pomeron structure function
parametrizations: one assumes that the pomeron is made of only quarks
(lower line), the other assumes that the pomeron has a dominant
hard gluon structure function (upper solid line). The hard quark
line underestimates the data by almost an order of magnitude, which cannot be
explained by the uncertainties in the flux assumed. 
ZEUS has done also a combined fit to these data and to the published $\ftwod$
1993 results \cite{zeusf2d93}, obtaining a gluon component of $\simeq 90\%$ at
$Q^2 = 4~\gevtwo$.

\section{Hard Diffraction at the Tevatron}

\begin{figure}
\mbox{
\psfig{figure=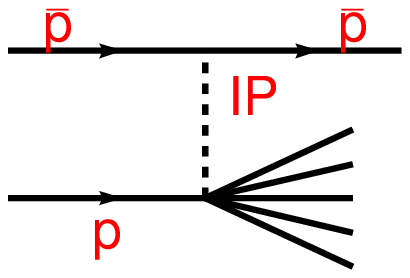,height=6cm}
\hspace{-1cm}
\psfig{figure=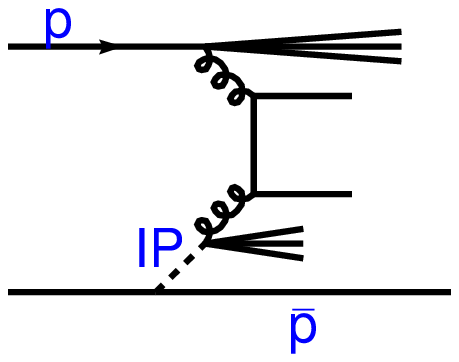,height=6cm}
\hspace{-1cm}
\psfig{figure=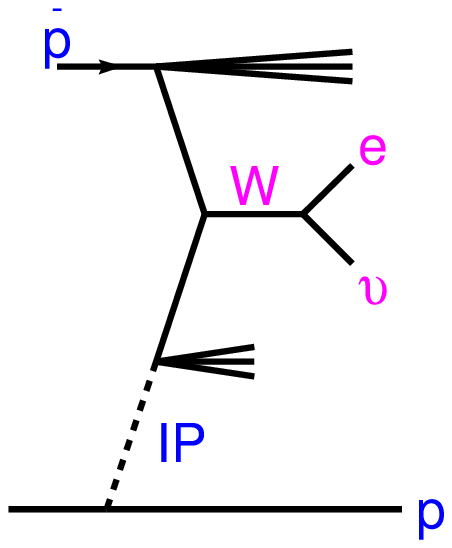,height=6cm}
}
\vspace{-2cm}
\caption{
Feynman diagram for single diffractive proton dissociation at Tevatron
(left) and in particular for dijet production (middle) and for $W$ production
(right).
\label{fig:ppdiagr}}
\end{figure}

Hard single diffraction in $p \bar p$ collisions at the Tevatron 
is another process that can be used to investigate the pomeron structure.
The process is the pomeron exchange reaction in the $t$ channel, 
in which the $\bar p$ ($p$) remains intact and
escapes down the beampipe, while the    
$p$ ($\bar p$) dissociates into a system separated in rapidity 
from the rest of the hadronic final state (see fig.~\ref{fig:ppdiagr}). 
The events are then characterized by a rapidity gap on one side.

These kind of events have been observed by both the 
CDF and D0 Collaborations in 
their 2-jet with high $E_T$ samples \cite{cdfdijet,d0dijet}. 
CDF has also observed hard diffraction
with production of a $W$ \cite{cdfwdiff}. 
These two processes give complementary information
on the pomeron structure:
the dijet sample gives information on the gluon content of the pomeron
(middle diagram in fig.~\ref{fig:ppdiagr}), while  the $W$ production 
is sensitive to the quark content in the pomeron 
(right diagram in fig.~\ref{fig:ppdiagr}).

\begin{figure}
\begin{center}
\mbox{
\psfig{figure=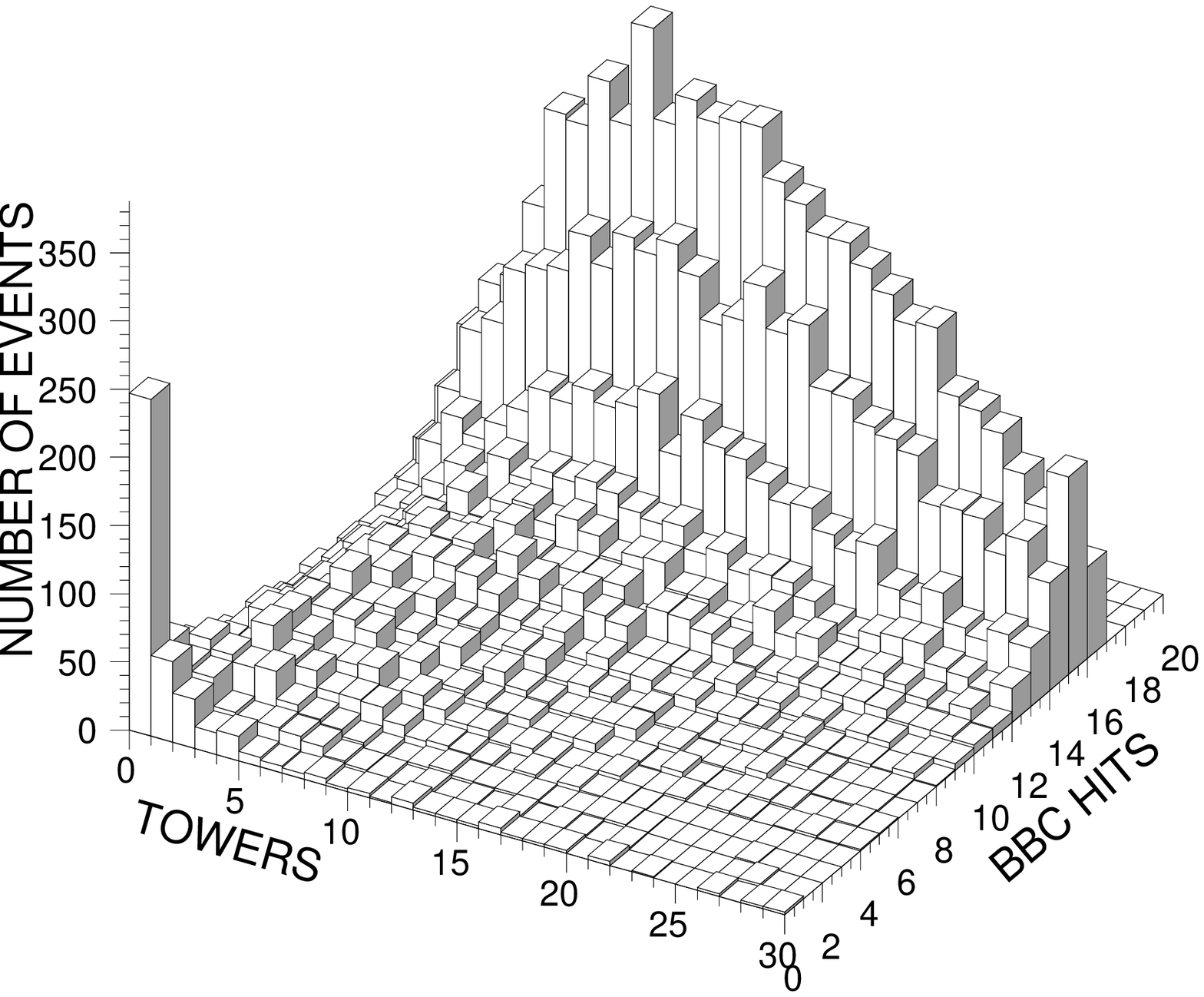,height=6cm}
\psfig{figure=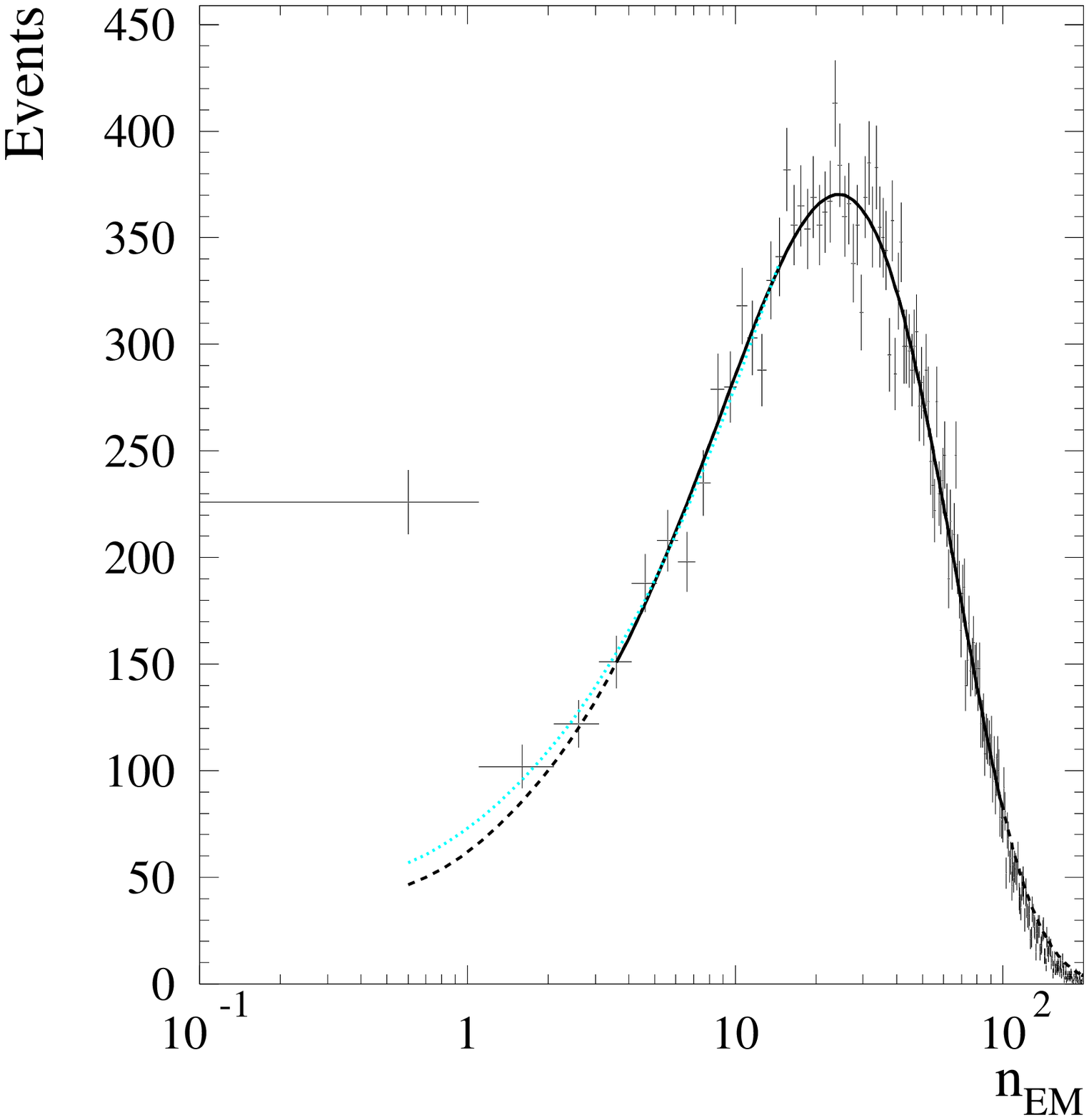,height=6cm}
}
\end{center}
\caption{
On the left: Correlation between the forward calorimeter tower multiplicity
 and the number of
hits in the BBC counter for the dijet CDF sample. On the right: multiplicity
measured in the D0 electromagnetic calorimeter for the dijet sample; the lines
represent two types of negative binomial fits to the non-diffractive part.
\label{fig:ppmult}}
\end{figure}

CDF and D0 search for single diffractive events by looking at the
measured multiplicity distributions. CDF looks, in the region opposite
the dijet system, for correlations
between the multiplicity measured in the forward part of the calorimeter 
($2.4<|\eta|<4.2$) and
the number of hits measured in a scintillator counter close to the beampipe,
the BBC ($3.2<|\eta|<5.9$). This correlation is shown for the dijet
samples with $E_T^{jet}>20$ GeV, $3.5>|\eta_{jet}|>1.8$ in 
fig.~\ref{fig:ppmult}. A clear peak at zero multiplicity in both parts
of the CDF detector can be seen and ascribed to single diffractive
events. By parametrizing the non-diffractive background using the shape from
the higher multiplicity part, the single diffractive signal can be extracted.
The ratio $R_{GJJ}= \sigma(Diffractive~Dijet)/\sigma(Dijet)$ is measured,
as in this ratio most of the systematic effects cancel. 
The ratio was corrected
for acceptance using a Monte Carlo based on a factorizable model and with
a hard gluon structure function, obtaining
\begin{equation}
R_{GJJ}= [0.75 \pm 0.05 (stat) \pm 0.09 (syst)] \% ~({\mathrm CDF~prel.}).
\end{equation}

D0 looked for single diffractive events in the dijet sample 
at $\sqrt s =1800~\gev$ ($E_T^{jet}>12$ GeV, $|\eta_{jet}|>1.6$), studying
the multiplicity in the electromagnetic calorimeter region ($2<|\eta|< 4.1$)
opposite the dijet system.
A peak at zero multiplicity can be seen in fig.~\ref{fig:ppmult},
consistent with a diffractive signal.
A negative binomial (NB) distribution was fit to the distribution,
excluding the very low multiplicity bins, 
to estimate the non-diffractive background. 
The uncorrected diffractive
signal, obtained from the excess over the NB fit extrapolated to zero 
multiplicity, is found to be:
\begin{equation}
R_{GJJ}=[ 0.67 \pm  0.05(stat+syst)]\%  ~({\mathrm D0~prel.}). 
\end{equation}

$W$ production in diffractive events at the Tevatron has been 
predicted~\cite{ingbruni}
and searched for by the CDF Collaboration. The sample of events currently
available is still
too small to look for multiplicity distributions. CDF exploits then certain
correlations which are present in these events 
(see fig.~\ref{fig:ppdiagr} right). Suppose that the $\bar p$
dissociates, giving a $W$ in the final state and the $p$ remains intact.
The events are then characterized by
a $W$ in the final state, which is detected in the analysis 
by the presence of a high $p_T$
electron (or positron) and a missing transverse momentum; 
furthermore a rapidity gap
on the opposite side of the electron or positron is present (this is called angle-gap
correlation). In addition, because of the high $W$ mass, the $W$ is likely
to be produced by a valence quark in the anti-proton and a valence quark in the pomeron
and, as the pomeron is quark-flavour symmetric, approximately two times more
electrons than positrons are expected in the final state (this is called charge-gap
correlation). 
Analogous correlations are found in the case in which the anti-proton
emits a pomeron.
In spite of the small signal (of the order of 20 events),
CDF can use these correlations to determine a 
ratio of diffractive $W$ production to non-diffractive $W$ production.
After correcting this measured ratio by acceptance using a factorizable Monte Carlo
model which assumes a hard quark structure function, CDF obtains:
\begin{equation}
R_W =    [1.15 \pm 0.51(stat) \pm 0.20 (syst)] \%.
\end{equation}
The two measured ratios can be combined to give the relative quark and
gluon content in the pomeron. In this combination, many assumptions on
the absolute normalization, like the pomeron flux and the momentum
sum in the pomeron, cancel out. The gluon content obtained by CDF is
\begin{equation}
c_g = 0.7 \pm 0.2,
\end{equation}
in agreement with what is found by the HERA experiments.

CDF and D0 have also observed candidate events for hard double pomeron
exchange (DPE). 
The process is quite interesting as both the proton and
the antiproton emit a pomeron, which then interact giving two high $E_t$ jets
in the final state. The events are characterized by two jets in the central region 
of the detector, and two rapidity gaps with respect to the two beam lines.
It is therefore a hard pomeron-pomeron scattering process. D0 observes a signal 
in the multiplicity distributions in the two opposite forward parts of the calorimeter.
CDF looks at the proton 
 side for multiplicity (in the same way as described above
for single diffraction) while tagging the anti-proton on the other side with the
forward proton spectrometer, requiring that $x_L\simeq 1$. 
Both experiments find a ratio of hard double pomeron exchange events to
non diffractive events of the order of $10^{-6}$, but more studies are needed
to confirm that it is really DPE. 

The double pomeron exchange process has been studied in \cite{berera}, 
as factorization breaking
effects have been predicted especially in hadron-hadron collisions \cite{jcollins}.
% The reason
%is that the two hadrons can interact strongly before the hard scattering, 
%breaking the factorization..... 
However the cross section calculated for DPE in
this non factorizable model is a couple of order of magnitudes
 higher than the measured one.

\section{A factorization test}

We have seen in this report interesting results on single diffraction
from HERA and Tevatron and on the partonic structure of the pomeron.
One of the main questions is if the pomeron vertex is factorizable,
that is if the HERA and Tevatron data can be explained by a universal
pomeron, with a universal flux and structure function.

Recently some groups have fit the HERA $\ftwod$ data and extracted
a pomeron structure function. They have then predicted the ratio $R_W$
at the Tevatron, assuming factorization in hard scattering.
 The cross section for $W$ production can be written
as (see fig.~\ref{fig:ppdiagr} right):
\begin{equation}
\sigma_{diff}  =  \sum_{a,b} f_{\pom/p}(\xpom,\mu) \otimes
                        f_{a/\bar p}(x_a,\mu) \otimes f_{b/\pom}(x_b,\mu) 
                                    \otimes \tilde \sigma_{ab}, 
\end{equation}
where $f_{\pom/p}(\xpom,\mu)$ is the flux of pomerons from the proton,
$f_{a/\bar p}(x_a,\mu)$ is the distribution function of parton $a$ in the
anti-proton, $f_{b/\pom}(x_b,\mu)$ is the distribution function of parton $b$
in the pomeron, 
while $\tilde \sigma_{ab}$ is the cross section for the $W$ production
from quarks $a$ and $b$, which is well known.

The groups use sligthly different assumptions for the flux and the pomeron and proton
structure functions and their evolution in $Q^2$. Alvero et al.~\cite{alvero},
when they use a gluon dominated parametrization for the
$\pom$ structure function, obtain a prediction of $R_W = 9.5\%$ for 
$\xpom<0.1$. Kunszt and Stirling~\cite{kustirling}, using three different models for
the HERA data, obtain a value for
the ratio $R_W$ varying from $5\%$ to $7\%$ for $\xpom<0.1$. 
Goulianos~\cite{goul2}, with 
simplified assumptions, obtains a value
$R_W=6.7\%$. These values have to be compared with the
measured ratio by CDF, $R_W=[1.15 \pm 0.55]\%$: they all seem to overestimate
$R_W$. However there are large uncertainties in
the $\xpom$ range covered by the large rapidity gap at CDF and
hence in these predictions which depend on $\xpom$. 

One of the possible explanations for the discrepancy could
be that factorization does not hold. Another possibility \cite{soper} is that 
interactions of the spectator quarks in the $p \bar p$ collision
could destroy the rapidity gap, which
has then a survival probability of the order of $10\%$~\footnote{This
survival rapidity was first introduced in \cite{bjorken} for central
rapidity gap events and predicted to be of the order of $5-30\%$.}. The
observed $R_W$ ratio may thus be reduced by this survival probability factor.

\section{Conclusions}

The recent results from HERA, E665 and the Tevatron on diffraction have
stimulated a lot of interest in pomeron exchange, as it is the ideal
ground to study the interplay between the soft non perturbative
regime and the hard perturbative region. 
Light vector meson or inclusive diffractive photoproduction
are well described by soft pomeron phenomenology. However,
as soon as an hard scale is involved in the process, for instance the mass
of the charm in the $J/\psi$ photoproduction or $Q^2$ in
inclusive photon dissociation, we observe a steep dependence of the
cross section with the energy, suggesting that perturbative QCD may
play a role.  

While pomeron exchange is well described in terms of Regge phenomenology, 
there is no prediction of its partonic structure, which has therefore to
be inferred from the experimental measurements. Results from the
diffractive structure function at HERA and from dijet and $W$
production at the Tevatron show that the pomeron is dominantly made
of gluons. Using these data, a test of the factorization of the
pomeron vertex has been presented. 
 
The pomeron structure function as obtained by fitting the HERA data
can be used to make prediction for single diffractive processes 
at future colliders. According to \cite{kusthiggs}, for example, 
$5-15\%$ of Higgs are produced in single diffractive processes at LHC. 
Therefore it will be important to look for rapidity gap events
also at future colliders.

\section*{Acknowledgments}

I would like to thank the organizers of the symposium for the invitation
to give this talk and for a nice conference. I would like also to thank the
many people who helped me
in preparing the plots: K.~Goulianos and P.~Melese (CDF); A.~Brandt and
T.~Taylor (D0); H.~Schellman (E665); J.~Dainton and P.~Newman (H1);
H.~Beier, G.~Briskin, N.~Cartiglia, J.~Puga, J.~Terron and S.M.~Wang (ZEUS). Finally special
thanks go to  M.~Arneodo, S.~Bhadra and R.~Nania who have assisted
me while preparing the talk and for a critical reading of this manuscript.

\end{document}